# Theory of Viscosity of Confined Fluids in Small / Nano Systems (Theory of Interfacial Viscosity)

**B. Mirzayi, G.A. Mansoori** [(*)] and **M. Vafaie-Sefti**

Departments of BioEngineering, Chemical Engineering & Physics
UNIVERSITY OF ILLINOIS AT CHICAGO
851 S. Morgan St. (M/C 063) Chicago, IL 60607-7052, USA
(*Dated: Thursday June 12, 2008*)

## Abstract

In this paper we present the molecular theory of viscosity of confined fluids in small / nano systems. This theory is also applicable to the interfacial viscosity. The basis of this research work is the Enskog's kinetic theory and the Boussinesq constitutive equation. The Enskog's kinetic theory is first transformed into a two-dimensional form. Then the potential energy (collisional transfer) part of the flux vector, and as a result, the contribution to the surface pressure tensor due to collisional transfer, is derived. Then the kinetic energy part of the flux vector and consequently the contribution to the surface pressure tensor due to flow of molecules is obtained.

The microscopic expression of total surface pressure tensor is obtained by adding of the potential energy (collisional transfer) part and the kinetic energy contribution. Then the expression of interfacial shear and dilatational viscosities are concluded by the comparison of corresponding terms of the two microscopic and macroscopic surface pressure tensor equations. Finally the dimensionless (reduced) forms of interfacial shear viscosity, interfacial dilatational viscosity and the surface tension equations are derived and they are plotted versus the reduced superficial number density.

(*) *Corresponding author*: mansoori@uic.edu

## Introduction

The behavior of small / nano systems like the interfacial region between two phases (fluid-solid or fluid-fluid) are of importance in the emerging field of nanotechnology, flow through porous media and in multiphase systems of scientific and industrial interest. The behavior of a confined fluid in a nano-space is different from those in macroscopic systems[1,2]. One of the properties of such small fluid systems which need understanding and formulation of its behavior, is the viscosity of confined fluids in small / nano systems.

One good example of a confined small fluid system property is the interfacial viscosity. The interfacial viscosity (IFV) is the two-dimensional equivalent to the bulk viscosity; it is the momentum transfer perpendicularly to the direction of motion on the interface. Interfacial viscosity is related to the interactions of molecules located at the interface with one another [3-5]. Molecular theory of several well-known bulk properties of fluids such as bulk viscosity, thermal conductivity and diffusion coefficient are already developed [6,7]. For the interface between two partially miscible fluids as well as between a fluid and a solid phase various researchers have investigated the molecular theories of surface tension, interfacial tension and wetability[8-11]. In the present report we introduce the molecular theory of interfacial viscosity.

## The Molecular Theory Interfacial Viscosity

The aim of this work is to develop the molecular expression of the two-dimensional viscosity (interfacial viscosity or IFV for short). Thus we perform all of the formulations in the two-dimensional system. The basis of this work is the Enskog's kinetic theory[12] and the Boussinesq constitutive equation[13].

The Enskog's theory[12] is the robust kinetic theory to describe the transport properties (such as viscosity, diffusion coefficient and thermal conductivity) in bulk dense gas systems. This theory first was derived for dilute gases in which it was assumed that, (i) there were only two-body collisions between molecules of the fluid, and (ii) the molecular diameter was small compared with the average distance between the molecules. These two assumptions are valid only in dilute gases. In dense gas and liquid systems, however, these assumptions will not be valid.

Enskog later developed a kinetic theory of dense gas systems[12]. For this model also only two-body collisions between molecules of the fluid are considered. By considering two-body collisions and by taking into account the finite size of the molecules Enskog was able to graft a theory of dense gas systems onto the dilute gas theory. For dense gas systems there are two effects which become important because molecules have finite volumes and diameters[12]: (i) *"Collisional transform" of momentum and energy* – when two molecules undergo a collision, there is an instantaneous transport of energy and momentum from the center of one molecule to the center of the other. (ii) *Change in the* *number of collisions per unit time* - the frequency of collisions is increased because molecular diameter is not negligibly small compared with the average distance between the molecules, and the frequency of collisions is decreased because the molecules are close enough to shield one another from oncoming molecules. This effect is described by Enskog scaling factor (Y). The factor Y is the pair correlation factor, which is intimately related to, the configurational properties, including the compressibility factor, as will be shown later.

Since the objective of the present report is the derivation of the molecular expression for the interfacial viscosity from the Enskog kinetic theory, this theory must be first transformed into a two-dimensional form. In the two following sections at first using the Enskog kinetic theory the potential energy (collisional transfer) part of the flux vector, as a result the contribution to the surface pressure tensor due to collisional transfer, is derived. Then the kinetic energy part of the flux vector and consequently the contribution to the surface pressure tensor due to flow of molecules is obtained.

## Potential energy (collisional transfer)

Let us consider a fluid composed of hard-sphere non-reacting and non-associating molecules of diameter $\sigma$ confined in the interface between two phases with infinite surfaces but with nano thickness of the interface. In this situation we assume the molecules can only move and collide in two dimensions. In discussing a two-dimensional (surface) collision between these molecules with velocities $\vec{\boldsymbol{v}}^s$, $\vec{\boldsymbol{v}}_1^s$ and centers O, $O_1$, let us consider the collisional transfer of a molecular property $\vec{\psi}^s$ across an element of line d$l$ at the point $\vec{\mathbf{r}}^s$. We use superscript "s" to denote the two-dimensional nature of variables and properties. The line element d$l$ is supposed to have a positive and a negative side and $\vec{\mathbf{n}}$ is the unit vector drawn normal to the element in the direction from the negative to the positive side. In this situation the product $\vec{\mathbf{i}}.\vec{\mathbf{n}}$ is positive (since $\vec{\mathbf{i}}$ is the unit vector in the direction of the first molecule's center from the second molecule). According to the Enskog kinetic theory the probable number of collisions between molecules with velocities $\vec{\boldsymbol{v}}^s$, $\vec{\boldsymbol{v}}_1^s$ (of which the first lies on the positive side of d$l$ and the second on the negative side) per unit time, where $\vec{\boldsymbol{v}}^s$, $\vec{\boldsymbol{v}}_1^s$ and $\vec{\mathbf{i}}$ lie in the ranges d$\vec{\boldsymbol{v}}^s$, d$\vec{\boldsymbol{v}}_1^s$ and d$\vec{\mathbf{i}}$ is[14]:

$$Y^s(\vec{\mathbf{r}}^s)(\vec{\psi}'^s - \vec{\psi}^s)f^s\left[\vec{\mathbf{r}}^s + \frac{1}{2}(\sigma\vec{\mathbf{i}})\right]f_1^s\left[\vec{\mathbf{r}}^s - \frac{1}{2}(\sigma\vec{\mathbf{i}})\right]\sigma^2(\vec{\mathbf{g}}^s.\vec{\mathbf{i}})(\vec{\mathbf{i}}.\vec{\mathbf{n}})d\vec{\mathbf{i}} \; d\vec{\boldsymbol{v}}^s d\vec{\boldsymbol{v}}_1^s \; dl \qquad (1)$$

where $\vec{\boldsymbol{v}}^s$, $\vec{\boldsymbol{v}}_1^s$ are the velocities of the colliding molecules, $\vec{\mathbf{i}}$ is a unit vector in the direction of the line of centers of molecules, $Y^s(\mathbf{r}^s)$ is the Enskog scaling factor which is just the equilibrium pair correlation function in contact, $f$ is the one-particle Boltzmann distribution function, $\vec{\mathbf{r}}^s + \frac{1}{2}(\sigma\vec{\mathbf{i}})$ and $\vec{\mathbf{r}}^s - \frac{1}{2}(\sigma\vec{\mathbf{i}})$ are the centers of the two colliding molecules (both with the same diameter $\sigma$), $\vec{\mathbf{g}}^s \equiv \vec{\mathbf{v}}^s - \vec{\mathbf{v}}_1^s$, $\vec{\mathbf{n}}$ is the unit vector normal to $dl$ and subscript (and superscript) "$s$" denotes the two-dimensional nature of a variable.

For a molecular property denoted by $\bar{\psi}^s$, at each collision, there is a net transfer of $\bar{\psi}'^s - \bar{\psi}^s$ across $dl$, where $\bar{\psi}^s$ is the value of the property prior to the collision and $\bar{\psi}'^s$ is the value after the collision. Thus the total transfer of $\bar{\psi}^s$ per unit length, *i.e.* by dropping $dl$ from Eq.(1), by all collisions is[14]:

$$Y^s(\vec{\mathbf{r}}^s)(\bar{\psi}'^s - \bar{\psi}^s)f^s\left[\vec{\mathbf{r}}^s + \frac{1}{2}\left(\sigma\vec{\mathbf{i}}\right)\right]f_1^s\left[\vec{\mathbf{r}}^s - \frac{1}{2}\left(\sigma\vec{\mathbf{i}}\right)\right]\sigma^2(\vec{\mathbf{g}}^s.\vec{\mathbf{i}})(\vec{\mathbf{i}}.\vec{\mathbf{n}})d\vec{\mathbf{i}}\ d\vec{\mathbf{v}}^sd\vec{\mathbf{v}}_1^s \qquad (2)$$

By taking integral over variables $\vec{\mathbf{v}}^s$, $\vec{\mathbf{v}}_1^s$ and $\vec{\mathbf{i}}$ the total rate of transfer of $\bar{\psi}^s$ across $dl$ by collisions, per unit time and per unit length is:

$$Y^s(\vec{\mathbf{r}}^s)\sigma^2\iint\int(\bar{\psi}'^s - \bar{\psi}^s)f^s\left[\vec{\mathbf{r}}^s + \frac{1}{2}\left(\sigma\vec{\mathbf{i}}\right)\right]f_1^s\left[\vec{\mathbf{r}}^s - \frac{1}{2}\left(\sigma\vec{\mathbf{i}}\right)\right](\vec{\mathbf{g}}^s.\vec{\mathbf{i}})(\vec{\mathbf{i}}.\vec{\mathbf{n}})d\vec{\mathbf{i}}\ d\vec{\mathbf{v}}^sd\vec{\mathbf{v}}_1^s \qquad (3)$$

In the above expression the integrations are over all values of variables such that $\vec{\mathbf{g}}^s.\vec{\mathbf{i}}$ and $\vec{\mathbf{i}}.\vec{\mathbf{n}}$ are positive. In this expression let the variables $\vec{\mathbf{v}}^s$, $\vec{\mathbf{v}}_1^s$ be interchanged; this is equivalent to interchanging the roles of the two colliding particles. Thus $\vec{\mathbf{i}}$ is replaced by $-\vec{\mathbf{i}}$, $\vec{\mathbf{g}}^s$ by $-\vec{\mathbf{g}}^s$, and ($\bar{\psi}'^s - \bar{\psi}^s$) by $-(\bar{\psi}'^s - \bar{\psi}^s)$. On performing the interchange of variables, we obtain an expression identical in form with the original one, Eq. (3), (i.e. prior to replacing the parameters $\vec{\mathbf{i}}$ and $\vec{\mathbf{g}}^s$ with $-\vec{\mathbf{i}}$, and $-\mathbf{g}^s$ respectively). But the integration now taken over all values of the variables such that $\mathbf{g}^s.\vec{\mathbf{i}}$ is positive and $\vec{\mathbf{i}}.\vec{\mathbf{n}}$ is negative. Hence the rate transfer of property $\bar{\psi}^s$ may be written as[6,7,14]:

$$\frac{1}{2}Y^s(\vec{\mathbf{r}}^s)\sigma^2\iint\int(\bar{\psi}'^s - \bar{\psi}^s)f^s\left[\vec{\mathbf{r}} + \frac{1}{2}\left(\sigma\vec{\mathbf{i}}\right)\right]f_1^s\left[\vec{\mathbf{r}} - \frac{1}{2}\left(\sigma\vec{\mathbf{i}}\right)\right](\vec{\mathbf{g}}^s.\vec{\mathbf{i}})(\vec{\mathbf{i}}.\vec{\mathbf{n}})d\vec{\mathbf{i}}\ d\vec{\mathbf{v}}^sd\vec{\mathbf{v}}_1^s \qquad (4)$$

The factor $\frac{1}{2}$ in this equation is to count every molecular collision only once. This expression is the scalar product of $\vec{\mathbf{n}}$ and another vector which represents the contribution of collisions to vector of flow of $\bar{\psi}^s$ (i.e. $\vec{\mathbf{n}}.\vec{\mathbf{\Psi}}_\Phi^s$). Thus we conclude:

$$\vec{\mathbf{\Psi}}_\Phi^s = \frac{1}{2}Y^s(\vec{\mathbf{r}}^s)\sigma^2\iint\int(\bar{\psi}'^s - \bar{\psi}^s)f^s\left[\vec{\mathbf{r}} + \frac{1}{2}\left(\sigma\vec{\mathbf{i}}\right)\right]f_1^s\left[\vec{\mathbf{r}} - \frac{1}{2}\left(\sigma\vec{\mathbf{i}}\right)\right](\vec{\mathbf{g}}^s.\vec{\mathbf{i}})\vec{\mathbf{i}}\ d\vec{\mathbf{i}}\ d\vec{\mathbf{v}}^sd\vec{\mathbf{v}}_1^s \qquad (5)$$

To solve this equation the one-particle Boltzmann distribution functions $f^s$ and $f_1^s$ must be known, but the exact expressions of these functions are not available. Therefore the approximate values of these functions may be used through the mathematical perturbation theory. Using the Taylor series expansions of $f^s \cdot f_1^s$ with respect to position vector ($\vec{\mathbf{r}}^s$) we get the following truncated perturbation expansion

$$f^s f_1^s = f^{[0]^s} f_1^{[0]^s} + \frac{1}{2}\sigma \ \vec{\mathbf{i}}.f^{[0]^s} f_1^{[0]^s} \frac{\partial}{\partial \vec{\mathbf{r}}^s} \ln \frac{f^{[0]^s}}{f_1^{[0]^s}} \ . \tag{6}$$

In the above equations $f^{[0]^s}$ and $f_1^{[0]^s}$ are the two-dimensional Maxwell-Boltzmann unperturbed distribution functions. By inserting Eq. (6) in Eq. (5) we get,

$$\vec{\boldsymbol{\Psi}}_\Phi^s = \frac{1}{2}\sigma^2 Y^s(\vec{\mathbf{r}}^s) \iiint (\vec{\psi}'^s - \vec{\psi}^s) f^{[0]^s} f_1^{[0]^s} (\vec{\mathbf{g}}^s . \vec{\mathbf{i}}) \vec{\mathbf{i}} \ d\vec{\mathbf{i}} \ d\vec{\mathbf{v}}^s d\vec{\mathbf{v}}_1^s$$

$$+ \frac{1}{4}\sigma^3 Y^s(\vec{\mathbf{r}}^s) \iiint (\vec{\psi}'^s - \vec{\psi}^s) \left( \vec{\mathbf{i}}.f^{[0]^s} f_1^{[0]^s} \frac{\partial}{\partial \vec{\mathbf{r}}^s} \ln \frac{f^{[0]^s}}{f_1^{[0]^s}} \right) (\vec{\mathbf{g}}^s . \vec{\mathbf{i}}) \vec{\mathbf{i}} \ d\vec{\mathbf{i}} \ d\vec{\mathbf{v}}^s d\vec{\mathbf{v}}_1^s + \cdots \tag{7}$$

In the above equation $\vec{\boldsymbol{\Psi}}_\Phi^s$ is the *collisional transfer* (due to potential energy) part of the flux vector property $\vec{\psi}^s$.

Since the IFV depends on the momentum transfer (*i.e.* mass × velocity), in the above equations $\vec{\psi}^s$ is replaced by $m\vec{\mathbf{v}}^s$ (where $\vec{\psi}^s$, m and $\vec{\mathbf{v}}^s$ are the surface flux vector, the mass of fluid and the surface mass average velocity vector, respectively). From this substitution the contributions to the two-dimensional (surface) pressure tensor due to potential energy (collisional transfer) may be obtained. Thus by setting $\vec{\psi}^s = m\vec{\mathbf{v}}^s$ in Eq.(7) we get:

$$\vec{\mathbf{P}}_\Phi^s = \frac{1}{2}\sigma^2 Y^s(\vec{\mathbf{r}}^s) \iiint (m\vec{\mathbf{v}}'^s - m\vec{\mathbf{v}}^s) f^{[0]^s} f_1^{[0]^s} (\vec{\mathbf{g}}^s . \vec{\mathbf{i}}) \vec{\mathbf{i}} \ d\vec{\mathbf{i}} \ d\vec{\mathbf{v}}^s d\vec{\mathbf{v}}_1^s$$

$$+ \frac{1}{4}\sigma^3 Y^s(\vec{\mathbf{r}}^s) \iiint (m\vec{\mathbf{v}}'^s - m\vec{\mathbf{v}}^s) \left( \vec{\mathbf{i}}.f^{[0]^s} f_1^{[0]^s} \frac{\partial}{\partial \vec{\mathbf{r}}^s} \ln \frac{f^{[0]^s}}{f_1^{[0]^s}} \right) (\vec{\mathbf{g}}^s . \vec{\mathbf{i}}) \vec{\mathbf{i}} \ d\vec{\mathbf{i}} \ d\vec{\mathbf{v}}^s d\vec{\mathbf{v}}_1^s + \cdots \tag{8}$$

In this equation $\vec{\mathbf{P}}_\Phi^s$ is the collisional transfer (potential energy) contribution to the surface pressure tensor. The Enskog scaling factor $Y^s$ is the representation of the number of molecular collisions (or probability of collisions) on a two dimensional interfacial surface. In such a two dimensional collision the hard-sphere molecules behave like disks. For hard-sphere molecules with diameter $\sigma$, the Enskog scaling factor Y can be determined using a number of methods[15]. Two methods are reported here to compute the factor Y. The first one is a very simple geometrical method, which refers the computation of Y to the increase of collision frequency due to the fact that the gas particles, in the Enskog gas, occupy a volume larger than zero. The second method equates Y to the *pair* *correlation function*[15]. The following expression for factor Y is obtained from using the first method:

$$Y = \frac{\left[1 - \frac{11}{12}\pi n\sigma^3\right]}{\left[1 - \frac{4}{3}\pi n\sigma^3\right]} = \frac{\left[1 - \frac{11}{8}b_0/V\right]}{\left[1 - 2b_0/V\right]} \tag{9}$$

Where $b_0/V = 2/3\pi n\sigma^3$, $b_0$ is referred to co-volume of the molecules, V is the volume of the system containing molecules and n is number density (number of molecules per unit volume). The above equation is extended by Clausius and Boltzmann as the following[16,17]:

$$Y = 1 + 0.6250\,b_0/V + 0.28695(b_0/V)^2 + 0.1103(b_0/V)^3 + 0.00386(b_0/V)^4 + ... \tag{10}$$

Besides the compressibility factor is defined by the following equation[7]:

$$Z = 1 + b_0/V + 0.6250(b_0/V)^2 + 0.28695(b_0/V)^3 + 0.1103(b_0/V)^4$$
$$+ 0.00386(b_0/V)^5 + ... \tag{11}$$

Combining equations (10) and (11) one obtains:

$$Z = 1 + \frac{b_0}{V}Y \Rightarrow Y = \left(\frac{Z-1}{b_0}\right)V \tag{12}$$

However when the molecules behave as a hard-disk the coefficient in equations (10) and (11) are changed. Ree and Hoover[18] calculated these coefficients for hard-disk molecules. Considering these coefficients for hard-disk molecules and changing the three-dimensional variables into two-dimensional forms the following equations yield:

$$Y^s = 1 + 0.7820\,b_0^s/\tilde{A} + 0.53223(b_0^s/\tilde{A})^2 + 0.3338(b_0^s/\tilde{A})^3 + 0.1992(b_0^s/\tilde{A})^4 + ... \tag{13}$$

$$Z^s = 1 + b_0^s/\tilde{A} + 0.7820(b_0^s/\tilde{A})^2 + 0.53223(b_0^s/\tilde{A})^3 + 0.3338(b_0^s/\tilde{A})^4$$
$$+ 0.1992(b_0^s/\tilde{A})^5 + ... \tag{14}$$

Combining equations (13) and (14) the following equation is obtained:

$$Z^s = 1 + \frac{b_0^s}{\tilde{A}}Y^s \Rightarrow Y^s = \left(\frac{Z^s - 1}{b_0^s}\right)\tilde{A} \tag{15}$$

In the above equations, $Y^s$ and $Z^s$ are the two dimensional Enskog scaling factor and compressibility factor, and

$$b_0^s/\tilde{A} = \frac{\pi\sigma^2 n^s}{2} \tag{16}$$

where $b_0^s$ is referred to co-area of the molecules[19-21], $\widetilde{A}$ is the area of the system containing molecules and $n^s$ is surface number density (number of molecules per unit area).

The function $f^{[0]s}$ in Eq. (8) is the two-dimensional Maxwell-Boltzmann unperturbed distribution function[22]:

$$f^{[0]s}(\vec{v}^s) = n^s \left(\frac{m}{2\pi kT}\right) e^{-m(\vec{v}^s - \vec{v}_0^s)^2/2kT} \tag{17}$$

In this equation $\vec{v}^s$ is the linear surface velocity vector and $\vec{v}_0^s$ is the mass average surface velocity vector.

The first integral in Eq (8) is as following:

$$I_1^s = \frac{1}{2}\sigma^2 Y^s(\vec{r}^s) \int\int\int (m\vec{v}'^s - m\vec{v}^s)\, f^{[0]s} f_1^{[0]s}\,(\vec{g}^s.\vec{i})\,\vec{i}\;d\vec{i}\;d\vec{v}^s d\vec{v}_1^s \tag{18}$$

For evaluating this integral the two following equations must be used [6,14]:

$$\int (\vec{v}'^s - \vec{v}^s)\;(\vec{g}^s.\vec{i})\,\vec{i}\;d\vec{i} = \int \vec{i}\,\vec{i}\,(\vec{g}^s.\vec{i})^2\,d\vec{i} = \frac{1}{8}\pi\left(2\,\vec{g}^s\vec{g}^s + g^{s2}\,\vec{I}_s\right) \tag{19}$$

$$\int\int (2\,\vec{g}^s\vec{g}^s + g^{s2}\,\vec{I}_s)d\vec{v}^s d\vec{v}_1^s = \left(2\,\overline{\vec{V}^s\vec{V}^s} + \overline{V^{s2}}\,\vec{I}_s\right) \tag{20}$$

Using the Eqs. (19) and (20) Eq. (18) yields:

$$I_1^s = \frac{1}{2}\sigma^2 Y^s(\vec{r}^s) \int\int\int (m\vec{v}'^s - m\vec{v}^s)\, f^{[0]s} f_1^{[0]s}\,(\vec{g}^s.\vec{i})\,\vec{i}\;d\vec{i}\;d\vec{v}^s d\vec{v}_1^s$$

$$= \frac{1}{16}(\pi m\sigma^2 Y^s) \int\int f^{[0]s} f_1^{[0]s}\,(2\,\vec{g}^s\vec{g}^s + g^{s2}\,\vec{I}_s)d\vec{v}^s d\vec{v}_1^s = \frac{1}{8}(\pi\,n^{s2}m\sigma^2 Y^s)\left(2\,\overline{\vec{V}^s\vec{V}^s} + \overline{V^{s2}}\,\vec{I}_s\right) \tag{21}$$

$$= \frac{1}{8}(\pi\,n^{s2}m\sigma^2 Y^s)\left(\vec{P}_K^s/n^s m + 3\,n^s kT\,\vec{I}_s/n^s m\right)$$

In above equation $\vec{V}^s = \vec{v}^s - \vec{v}_0^s$ is the peculiar surface velocity vector, $\overline{V^{s2}} = 3kT/m$ is the mean peculiar surface velocity and $\vec{P}_K^s = 2n^s m\,\overline{\vec{V}^s\vec{V}^s}$ is the kinetic contribution to the surface pressure tensor which will be obtained in the second subsection.

The second integral in Eq. (8) through Eq. (19) is written as:

$$I_2^s = \frac{1}{4}\sigma^3 Y^s(\vec{r}^s)\iiint(m\vec{v}'^s - m\vec{v}^s)\left(\vec{i}.f^{[0]s}f_1^{[0]s}\frac{\partial}{\partial\vec{r}^s}\ln\frac{f^{[0]s}}{f_1^{[0]s}}\right)(\vec{g}^s.\vec{i})\,\vec{i}\,d\vec{i}\,d\vec{v}^s d\vec{v}_1^s =$$

$$\frac{1}{4}\sigma^3 Y^s(\vec{r}^s)\,m\iiint\vec{i}\,\vec{i}\left(\vec{i}.f^{[0]s}f_1^{[0]s}\frac{\partial}{\partial\vec{r}^s}\ln\frac{f^{[0]s}}{f_1^{[0]s}}\right)(\vec{g}^s.\vec{i})^2\,d\vec{i}\,d\vec{v}^s d\vec{v}_1^s \qquad (22)$$

To evaluate this integral the following relation must be utilized [6,14]:

$$\int\vec{i}\,\vec{i}\,(\vec{i}.\vec{U})\,(\vec{g}^s.\vec{i})^2\,d\vec{i} = \frac{4}{15}\left[\vec{U}.\vec{g}^s\left(\vec{g}^s\vec{g}^s + g^{s2}\vec{I}_s\right)/g^s + g^s\left(\vec{U}\,\vec{g}^s + \vec{g}^s\,\vec{U}\right)\right] \qquad (23)$$

where $\vec{U}$ is the any arbitrary vector.

By taking $\frac{\partial}{\partial\vec{r}^s}\ln\frac{f^{[0]s}}{f_1^{[0]s}}$ instead of $\vec{U}$ in Eq (23) and using of this equation, the Eq. (22)

may be written as:

$$I_2^s = \frac{1}{4}\sigma^3 Y^s(\vec{r}^s)\,m\iiint\vec{i}\,\vec{i}\left(\vec{i}.f^{[0]s}f_1^{[0]s}\frac{\partial}{\partial\vec{r}^s}\ln\frac{f^{[0]s}}{f_1^{[0]s}}\right)(\vec{g}^s.\vec{i})^2\,d\vec{i}\,d\vec{v}^s d\vec{v}_1^s$$

$$= \frac{1}{15}(m\sigma^3 Y^s)\iint f^{[0]s}f_1^{[0]s}\left[\begin{array}{l}\left[\vec{g}^s.\frac{\partial}{\partial\vec{r}^s}\ln\frac{f^{[0]s}}{f_1^{[0]s}}\right]\frac{(\vec{g}^s\vec{g}^s + g^{s2}\,\vec{I}_s)}{g^s} \\ + g^s\left(\vec{g}^s\frac{\partial}{\partial\vec{r}^s}\ln\frac{f^{[0]s}}{f_1^{[0]s}} + \frac{\partial}{\partial\vec{r}^s}\ln\frac{f^{[0]s}}{f_1^{[0]s}}\vec{g}^s\right)\end{array}\right]d\vec{v}^s d\vec{v}_1^s \qquad (24)$$

In above equation for acquiring the $\frac{\partial}{\partial\vec{r}^s}\ln\frac{f^{[0]s}}{f_1^{[0]s}}$, Eq. (17) must be used. From Eq (17)

the ratio $\frac{f^{[0]s}}{f_1^{[0]s}}$ and thus $\ln\frac{f^{[0]s}}{f_1^{[0]s}}$ can be obtained. By taking the first derivation of $\ln\frac{f^{[0]s}}{f_1^{[0]s}}$

function the following expression may be concluded:

$$\frac{\partial}{\partial\vec{r}^s}\ln\frac{f^{[0]s}}{f_1^{[0]s}} = \frac{m}{2kT^2}(V^{s2} - V_1^{s2})\vec{\nabla}_s T + \frac{m}{kT}(\vec{V}^s - \vec{V}_1^s).\vec{\nabla}_s\vec{v}_0^s \qquad (25)$$

The terms involving $\vec{\nabla}_s T$ in Eq. (25) are odd functions of $\vec{V}^s$ and $\vec{V}_1^s$ and thus vanish on integration. The remaining terms are more conveniently evaluated by changing the variables of integration from $\vec{v}^s$ and $\vec{v}_1^s$ to $\vec{G}_0^s = (\vec{V}^s + \vec{V}_1^s)/2$ and $\vec{g}^s$ $(\vec{v}_1^s - \vec{v}^s)$, hence the integral is equal to:

$$I_2^s = \frac{m^2 \sigma^3 Y^s}{15 \, kT} \left( \frac{n^s \, m}{2\pi kT} \right)^2 \int\int \exp\left[ -\frac{m}{kT} \left( G_0^{s\,2} + \frac{1}{4} g^{s\,2} \right) \right]$$

$$\times \left\{ \bar{\nabla}_s \vec{v}_0^s : \frac{\vec{g}^s \vec{g}^s (\vec{g}^s \vec{g}^s + g^{s\,2} \, \vec{I}_s)}{g^s} + g^s \left[ \left( \bar{\nabla}_s \vec{v}_0^s . \vec{g}^s \right) \vec{g}^s + \vec{g}^s \left( \bar{\nabla}_s \vec{v}_0^s . \vec{g}^s \right) \right] \right\} d\vec{G}_0^s d\vec{g}^s \qquad (26)$$

Performing the integration indicated in Eq (26) yields:

$$I_2^s = -\frac{5 \, n^{s\,2} \sigma^3 Y^s}{8} (\pi m kT)^{0.5} \left[ \frac{4}{5} \, \vec{D}_s + \frac{4}{5} \, \vec{I}_s \bar{\nabla}_s . \vec{v}_0^s \right] \qquad (27)$$

In this equation $\vec{D}_s$ is the surface rate of deformation tensor defined as:

$$\vec{D}_s = \frac{1}{2} \left[ \left( \bar{\nabla}_s \vec{v}_0^s \right) . \vec{I}_s + \vec{I}_s . \left( \bar{\nabla}_s \vec{v}_0^s \right)^T \right] \qquad (28)$$

where superscript $T$, $\vec{I}_s$, $\bar{\nabla}_s$ and $\vec{v}_0^s$ are the transpose, surface unit tensor, surface gradient operator and mass average surface velocity vector, respectively.

The total contribution to momentum flux due to potential energy (collisional transfer) is the sum of the expressions in Eqs. (21) and (27). The following expression is obtained from this summation:

$$\vec{P}_\Phi^s = I_1^s + I_2^s = \frac{1}{8} (\pi \, n^{s\,2} m \sigma^2 Y^s) \left( \vec{P}_K^s / n^s m + 3 n^s kT \, \vec{I}_s / n^s m \right)$$

$$- \frac{5 \, n^{s\,2} \sigma^3 Y^s}{8} (\pi m kT)^{0.5} \left[ \frac{4}{5} \, \vec{D}_s + \frac{4}{5} \, \vec{I}_s \bar{\nabla}_s . \vec{v}_0^s \right] \qquad (29)$$

## Kinetic energy (Molecular flow)

In this section the derivation of kinetic energy part of the flux vector property and consequently the contribution to the surface pressure tensor due to flow of molecules is discussed. $\vec{\Psi}_K^s$, the kinetic energy part of the flux vector property $\bar{\psi}^s$, is defined by[7]:

$$\vec{\Psi}_K^s = \int \bar{\psi}^s \, f^s \, \vec{V}^s \, d\vec{v}^s \qquad (30)$$

In this equation $f^s$ is the two-dimensional Boltzmann distribution function, which is derived by solving the Boltzmann equation as[7]:

$$f^s (\vec{r}^s, \vec{v}^s, t) = f^{[0]s} (\vec{v}^s) \left[ 1 + \phi^s (\vec{r}^s, \vec{v}^s, t) \right] \qquad (31)$$

In above equation $\phi^s$ is the two-dimensional perturbation function and defined as [23]:

$$\phi^s = -\frac{1}{Y^s}\left(1+\frac{3}{8}\pi n\sigma^2 Y^s\right)\left(\vec{A}\cdot\frac{\partial\ln T}{\partial\vec{r}^s}\right) - \frac{1}{Y^s}\left(1+\frac{1}{4}\pi n\sigma^2 Y^s\right)\left(\vec{B}:\vec{\nabla}\,\vec{v}_0^s\right) \tag{32}$$

where parameters $\vec{A}$ and $\vec{B}$ are defined as [7]:

$$\vec{A} = f^{[0]s}\left(\frac{5}{2}-W^{s\,2}\right)\vec{V}^s \tag{33}$$

$$\vec{B} = -2f^{[0]s}\left(\vec{W}^s\vec{W}^s - \frac{1}{3}W^{s\,2}\,\vec{I}_s\right) \tag{34}$$

and

$$\vec{W}^s = \sqrt{m/2kT}\;\vec{V}^s \tag{35}$$

is the reduced velocity vector.

By setting $\vec{\psi}^s = m\vec{v}^s$ in Eq. (30) and using the two-dimensional Boltzmann distribution function (Eq. 31) the kinetic contribution to the surface pressure tensor is obtained. This substitution leads to $\vec{\Psi}_K^s$ changing to $\vec{P}_K^s$. The result is [14,23]:

$$\vec{P}_K^s = n^s kT\,\vec{I}_s - \frac{2}{Y^s}\left(1+\frac{1}{4}\pi n^s\sigma^2 Y^s\right)\eta^{0s}\,\vec{D}_s \tag{36}$$

In the above equation $\vec{P}_K^s$ is the kinetic contribution to the surface pressure tensor and

$$\eta^{0s} = \frac{1.022}{2\sigma}\sqrt{\frac{mkT}{\pi}} \tag{37}$$

is the shear viscosity of hard-disk dilute gas [24,25].

In the following section the microscopic expression of total surface pressure tensor is obtained by adding of the potential energy (collisional transfer) part, Eq. (29), and the kinetic energy contribution, Eq. (36). Then the macroscopic relation of total surface pressure tensor is derived. Finally the expression of interfacial shear and dilatational viscosities are concluded by the comparison of corresponding terms of the two microscopic and macroscopic surface pressure tensor equations.

## Surface pressure tensor plus interfacial shear and dilatational viscosities

By substituting equation (36) into Eq.(29) the potential energy term of surface pressure tensor will be as follows:

$$
\begin{aligned}
\vec{P}_\Phi^s &= \frac{1}{2}(n^s kT)(\pi n^s \sigma^2 Y^s)\vec{I}_s - \frac{1}{4}(\pi n^s \sigma^2 Y^s)\left(1 + \frac{1}{4}\pi n^s \sigma^2 Y^s\right)\eta^{0s}\,\mathbf{D}_s \\
&\quad - \frac{5 n^{s2}\sigma^3 Y^s}{8}(\pi m kT)^{0.5}\left[\frac{4}{5}\vec{D}_s + \frac{4}{5}\vec{I}_s\vec{\nabla}_s\cdot\vec{v}_0^s\right] \\
&= \left[\frac{1}{2}(n^s kT)(\pi n^s \sigma^2 Y^s) - \frac{n^{s2}\sigma^3 Y^s}{2}(\pi m kT)^{0.5}\vec{\nabla}_s\cdot\vec{v}_0^s\right]\vec{I}_s \\
&\quad - \left[\frac{1}{4}(\pi n^s \sigma^2 Y^s)\left(1 + \frac{1}{4}\pi n^s \sigma^2 Y^s\right)\eta^{0s} + \frac{n^{s2}\sigma^3 Y^s}{2}(\pi m kT)^{0.5}\right]\vec{D}_s
\end{aligned}
\tag{38}
$$

The total surface pressure tensor is the sum of the potential energy (collisional transfer) part, Eq. (29), and the kinetic energy contribution, Eq. (36). By adding these two terms we get the total surface pressure tensor expression as the following:

$$
\begin{aligned}
\vec{P}^s &= \vec{P}_K^s + \vec{P}_\Phi^s = \left[n^s kT + \frac{1}{2}(n^s kT)(\pi n^s \sigma^2 Y^s) - \frac{n^{s2}\sigma^3 Y^s}{2}(\pi m kT)^{0.5}\vec{\nabla}_s\cdot\vec{v}_0^s\right]\vec{I}_s \\
&\quad - \left[(\frac{1}{4}\pi n^s \sigma^2 Y^s + \frac{2}{Y^s})\left(1 + \frac{1}{4}\pi n^s \sigma^2 Y^s\right)\eta^{0s} + \frac{n^{s2}\sigma^3 Y^s}{2}(\pi m kT)^{0.5}\right]\vec{D}_s
\end{aligned}
\tag{39}
$$

In above equation $\vec{P}^s$ is the surface pressure tensor. This expression for surface pressure tensor is obtained from the two-dimensional (surface) discussion of kinetic theory and involves only molecular properties of molecules on the surface. In the continuation of this discussion we explain the macroscopic equation of surface pressure tensor and compare that with the microscopic equation of surface pressure tensor to obtain the molecular equations of interfacial shear and dilatational viscosities.

Boussinesq[13] postulated the existence of a 'surface viscosity' conceived as the two-dimensional equivalent of conventional three-dimensional viscosity possessed by bulk-fluid phases. According to fluid mechanics, under a dynamic condition the surface (two-dimensional) pressure tensor $\vec{P}^s$, is defined in terms of *isotropic* and *deviatoric* (viscous) parts[13]:

$$
\vec{P}^s = \gamma\vec{I}_s - \vec{\tau}^s
\tag{40}
$$

In this equation the first term is the isotropic part ($\gamma$ is the thermodynamic interfacial tension) and the second term is the deviatoric (viscous) part ($\vec{\tau}^s$ is the surface stress tensor).

To obtain an explicit expression for surface pressure tensor $\vec{P}^s$, a specific Boussinesq-Scriven[13,26], constitutive equation ,Eq. (40),  for the surface stress tensor $\vec{\tau}^s$, must be used:

$$\vec{\tau}^s = \left(\eta_d^s - \eta_{sh}^s\right)\left(\vec{\nabla}_s \cdot \vec{v}_0^s\right)\vec{I}_s + 2\,\eta_{sh}^s\,\vec{D}_s \tag{41}$$

which is the two-dimensional form of the stress tensor and is composed of two parts, a *shear* and a *dilatational* component.

Combination of Eqs.(39) and (40) yields:

$$\vec{P}^s = \gamma\vec{I}_s - \left(\eta_d^s - \eta_{sh}^s\right)\left(\vec{\nabla}_s \cdot \vec{v}_0^s\right)\vec{I}_s - 2\,\eta_{sh}^s\,\vec{D}_s = \left[\gamma - \left(\eta_d^s - \eta_{sh}^s\right)\left(\vec{\nabla}_s \cdot \vec{v}_0^s\right)\right]\vec{I}_s - 2\,\eta_{sh}^s\,\vec{D}_s \tag{42}$$

Comparison of the second terms of Eq. (41) with Eq. (38) gives the *interfacial shear viscosity* $\eta_{sh}^s$, as:

$$\eta_{sh}^s = \frac{1}{2}\left[\left(\frac{1}{4}\pi\,n^s\,\sigma^2 Y^s + \frac{2}{Y^s}\right)\left(1 + \frac{1}{4}\pi\,n^s\,\sigma^2 Y^s\right)\eta^{0s} + \frac{n^{s^2}\sigma^3 Y^s}{2}\left(\pi mkT\right)^{0.5}\right] \tag{43}$$

The *interfacial dilatational viscosity* $\eta_d^s$, is obtained by comparison of the coefficients of $\vec{\nabla}_s \cdot \vec{v}_0^s$ in first term of Eq. (41) with Eq.(38):

$$\eta_d^s = \eta_{sh}^s + \frac{n^{s^2}\sigma^3 Y^s}{2}\left(\pi mkT\right)^{0.5} \tag{44}$$

The interfacial shear viscosity $\eta_{sh}^s$, and the interfacial dilatational viscosity $\eta_d^s$, can be predicted by the determination of the parameters in equations (42) and (43).

The other expression that we can extract from the comparison of  Eqs. (38) and (41) is the surface tension equation. The first term of first bracket in Eq. (38) is equal to surface tension ($\gamma$) which appears in Eq. (41). The result of this comparison is:

$$\gamma = n^s kT + \frac{1}{2}\left(n^s\,kT\right)\left(\pi n^s\,\sigma^2 Y^s\right) \tag{45}$$

It is more convenient to use the dimensionless forms. Thus in next section the reduced (dimensionless) form of Eqs. (42) and (43) are derived.

By definition of the $\eta_{sh}^{s*} = \dfrac{\eta_{sh}^{s}}{\eta^{s0}}$, $\eta_{d}^{s*} = \dfrac{\eta_{d}^{s}}{\eta^{s0}}$ and $n^{s*} = n^{s} \sigma^{2}$ - where $\eta_{sh}^{s*}$, $\eta_{d}^{s*}$ and $n^{s*}$ are reduced interfacial shear viscosity, reduced interfacial dilatational viscosity and reduced superficial number density, respectively- the dimensionless form of Eqs. (42) and (43) can be derived. If two sides of Eq. (42) are divided by $\eta^{s0}$, the reduced interfacial shear viscosity, $\eta_{sh}^{s*}$ is obtained. The result is as the following:

$$\eta_{sh}^{s*} = \frac{1}{2}\left[ \left( \frac{1}{4}\pi n^{s*} Y^{s} + \frac{2}{Y^{s}} \right)\left( 1 + \frac{1}{4}\pi n^{s*} Y^{s} \right) + \frac{\pi \left( n^{s*} \right)^{2} Y^{s}}{1.022} \right] \qquad (46)$$

and the reduced interfacial dilatational viscosity, $\eta_{d}^{s*}$ is obtained from Eq.(43) by dividing it to $\eta^{s0}$ :

$$\eta_{d}^{s*} = \eta_{sh}^{s*} + \frac{\pi n^{s*} Y^{s}}{2.044} \qquad (47)$$

The reduced form of Eq. (44) is obtained by reduced surface tension , $\gamma^{*} = \dfrac{\gamma}{n^{s}kT}$ . To derive the reduced surface tension equation we divide the two sides of Eq. (44) to $n^{s}kT$ . The result is as the following:

$$\gamma^{*} = \frac{\gamma}{n^{s}kT} = 1 + \frac{1}{2}(\pi n^{s*} Y^{s}) \qquad (48)$$

In the above equation $\gamma^{*}$ is the reduced surface tension.

Considering the fact that $Y^{s}$ is function of $n^{s*}$ one may conclude that the shear and dilatational viscosities are functions of only $n^{s*}$ . In Figures 1 and 2 we report the dimensionless (reduced) viscosities and surface tension versus $n^{s*}$ , the reduced superficial number density.

The interfacial viscosity is the two-dimensional equivalent to the conventional three-dimensional bulk viscosity possessed by bulk fluid phases. Interfacial viscosity is related to the interactions of molecules located at interface with one another but the bulk viscosity is depends on the interaction of bulk molecules. The well-known *Naveir-Stokes* momentum transfer equation, in which the bulk viscosity is the main fluid property, represents the motion of bulk fluid phases however dynamic processes involving fluid interfaces which are represented by *Boussinesq-Scriven* constitutive equation depend on the interfacial viscosities.

In the Naveir-Stokes momentum transfer equation the three-dimensional form of Eqs. (35) and (36) are used (see Eqs (43) and (44) in Ref. 7). In other word, the two-dimensional properties such as interfacial viscosities, interfacial tension,… are replaced by the three-dimensional properties (i.e. bulk viscosity, thermodynamic pressure, …).

$$\vec{\tau} = \eta_d \left(\vec{\nabla} \cdot \vec{v}_0\right)\vec{I} + 2\,\eta_{sh}\vec{D} \tag{49}$$

$$\vec{P} = p\vec{I} - \eta_d \left(\nabla \cdot \vec{v}_0\right)\vec{I} - 2\eta_{sh}\vec{D} = \left[p - \eta_d \left(\vec{\nabla} \cdot \vec{v}_0\right)\right]\vec{I} - 2\,\eta_{sh}\vec{D} \tag{50}$$

The molecular expression of interfacial viscosities (Eqs. (37) and (38)), include the interfacial (two-dimensional) properties however the molecular equation of bulk viscosity (Eqs. (45) and (46)) 7 which is derived from the combination of three-dimensional kinetic theory with Naveir-Stokes equation consists the bulk (three-dimensional) molecular properties.

$$\eta_{sh} = \frac{1}{Y}\left(1 + \frac{4}{15}\pi n\sigma^3 Y\right)^2 \eta^0 + \frac{12}{45}n^2\sigma^4 Y\sqrt{\pi mkT} \tag{51}$$

$$\eta_d = \frac{4}{9}n^2\sigma^4 Y\sqrt{\pi mkT} \tag{52}$$

In above equations the three-dimensional parameters are, $\vec{\nabla}$: gradient operator, $\vec{\tau}$: stress tensor, $\eta^0$: dilute gas viscosity, $\eta_d$: bulk dilatational viscosity, $\eta_{sh}$: bulk shear viscosity, $\vec{v}_0$: mass average velocity vector, $\vec{D}$: rate of deformation tensor, $\vec{I}$: unit tensor, n: number density, $\vec{P}$: pressure tensor, p: thermodynamic pressure and Y: the Enskog scaling factor.

## Conclusions

In this paper we have developed the molecular theory of viscosity of confined fluids in small / nano systems. This theory is also the same as the interfacial viscosity. We have utilized the Enskog's kinetic theory and the Boussinesq constitutive equation as the basis of this research work. To achieve this the Enskog's kinetic theory equation is at first transformed into a two-dimensional format. Then the potential energy (collisional transfer) part of the flux vector, and as a result, the contribution to the surface pressure tensor due to collisional transfer, is derived. Then the kinetic energy part of the flux vector and consequently the contribution to the surface pressure tensor due to flow of molecules is obtained.

The microscopic expression of total surface pressure tensor is obtained by adding of the potential energy (collisional transfer) part and the kinetic energy contribution. Then the expression of interfacial shear and dilatational viscosities are concluded by the comparison of corresponding terms of the two microscopic and macroscopic surface pressure tensor equations. Finally the dimensionless (reduced) forms of interfacial shear viscosity, interfacial dilatational viscosity and the surface tension equations are derived and they are plotted versus the reduced superficial number density.

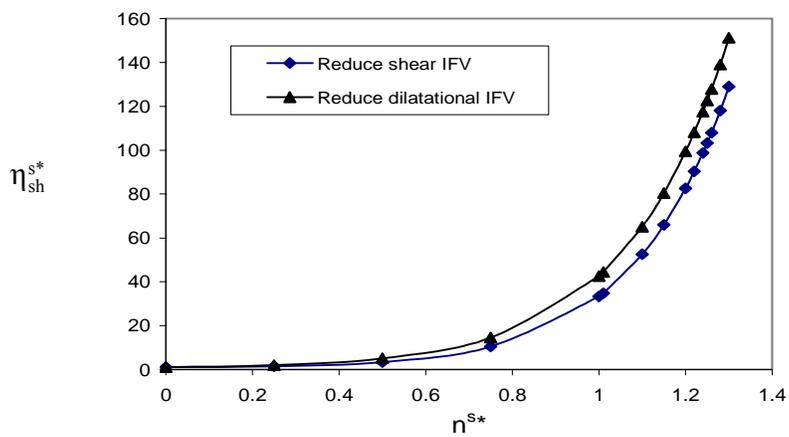

**Figure 1**: Reduced interfacial shear and dilatational viscosities vs. $n^{s*}$

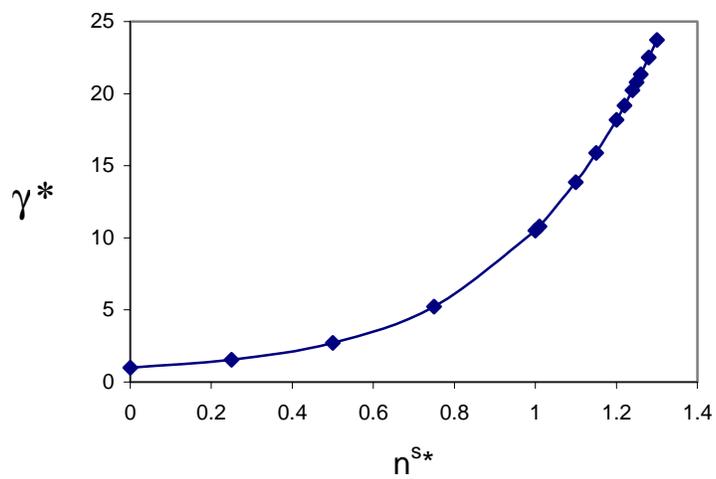

**Figure 2**: Reduced surface tension vs. $n^{s*}$